\documentclass[12pt]{extarticle}
\usepackage{tabularx} 
\usepackage{preamble}
\usepackage{graphicx}
\usepackage{packages}
\usepackage{breqn}
\usepackage{enumerate}
\usepackage[font=small,labelfont=bf]{caption}
\usepackage[left=30mm,right=30mm,top=30mm,bottom=30mm]{geometry}
\usepackage{pgfplots}
\usetikzlibrary{calc}
\usepackage{extarrows}
\usepackage{amsmath,amssymb}
\usepackage{amsfonts}
\usepackage{booktabs,subcaption,amsfonts,dcolumn}
\newcolumntype{d}[1]{D..{#1}}
\definecolor{refkey}{rgb}{0.9451,0.2706,0.4941}
\definecolor{labelkey}{rgb}{0.9451,0.2706,0.4941}
\DeclareFontFamily{U}{rcjhbltx}{}
\DeclareFontShape{U}{rcjhbltx}{m}{n}{<->rcjhbltx}{}
\DeclareSymbolFont{hebrewletters}{U}{rcjhbltx}{m}{n}
\DeclareMathSymbol{\tet}{\mathord}{hebrewletters}{84}
\DeclareMathSymbol{\pey}{\mathord}{hebrewletters}{112}

\def\z2{$\mathbb{Z}_2$}

\definecolor{darkgray}{rgb}{0.33, 0.33, 0.33}
\usepackage{braket}

\newcommand{\Tr}{{\mathrm{Tr}}}
\newcommand{\al}{\alpha'}

\newcommand{\be}{\begin{equation}}
\newcommand{\ba}{\begin{eqnarray}}
\newcommand{\ea}{\end{eqnarray}}
\newcommand{\ee}{\end{equation}}

\newcommand{\ap}{\alpha}

\newcommand{\la}{\langle}

\newcommand{\bea}{\begin{eqnarray}}
\newcommand{\eea}{\end{eqnarray}}
\newcommand{\bes}{\begin{equation*}}
\newcommand{\beas}{\begin{eqnarray*}}
\newcommand{\eeas}{\end{eqnarray*}}
\newcommand{\bas}{\begin{array*}}
\newcommand{\eas}{\end{array*}}
\newcommand{\ees}{\end{equation*}}
\newcommand{\nn}{\nonumber}

\newcommand{\ep}{\epsilon}

\newcommand{\ra}{\rangle}

\let\a=\alpha   \let\d=\delta \let\e=\epsilon   \let\h=\eta  \let\l=\lambda  
  \let\r=\rho  \let\t=\tau      \let\z=\zeta
          
\def\nn{\nonumber}
\def\inf{\infty}

\def\pa{\partial}

\usepackage{authblk}
\begin{document}

\title{\vspace{0.8cm}\makebox[\textwidth][c]{\Huge \bf Information metric on the boundary}\vspace{0.5cm}}
\author[1,2]{Kenta Suzuki}
\author[2]{Yu-ki Suzuki\footnote{yu-ki.suzuki@yukawa.kyoto-u.ac.jp}}
\author[2]{Takashi Tsuda}
\author[2]{Masataka~Watanabe\vspace{0.5cm}}
\affil[1]{{\small Department of Physics, Rikkyo University, Toshima, Tokyo 171-8501, Japan}}
\affil[2]{{\small Center for Gravitational Physics, Yukawa Institute for Theoretical Physics,\par
Kyoto University, Sakyo-ku, Kyoto 606-8502, Japan}}
\date{}
\pagenumbering{gobble}

\maketitle

\vspace*{-10.5cm}
\begin{flushright}
RUP-22-28
\\
YITP-22-156
\\
\vspace*{10cm}
\end{flushright}

\begin{abstract}

The information metric on the space of boundary coupling constants in two-dimensional conformal field theories is studied.
{Such a metric is related to the Casimir energy difference of the theory defined on an interval.}
We concretely compute the information metric on the boundary conformal manifold of free boson CFT as well as $SU(2)_k$ WZW theory, obtaining the result expected from the symmetry of the systems.
We also compute the information metric on the space of non-conformal boundary states produced by boundary mass perturbations in the theory of a real free scalar.
The holographic dual of the boundary information metric in the context of AdS$_3$/BCFT$_2$ is also discussed.
We argue that it corresponds to the area of the minimal cross section of the end-of-the-world brane connecting two boundaries of the asymptotic BCFTs.



\end{abstract}

\newpage
\pagenumbering{arabic}
\pagestyle{plain}

\tableofcontents

\section{Introduction}


As no actual experiments can be performed on a strictly infinite volume space, understanding of boundary conditions is a natural subject in physics.
One of the most well-studied classes of theories with boundaries is the boundary conformal field theory (BCFT) \cite{Cardy:1984bb, Cardy:1989ir}.
BCFT is a conformal field theory (CFT) defined on a space-time with boundaries, which are invariant under the maximal subgroup of the bulk conformal symmetry.
BCFTs are naturally useful in analyzing the long-wavelength limit of spin chains with a boundary or a defect \cite{Cardy:2004hm}, and in describing the worldsheet theory of the open string  \cite{Callan:1987px, Abouelsaood:1986gd, Polchinski:1987tu}.
Less trivially, they give a natural tensor factorization structure of the Hilbert space used to compute various quantum information theoretic quantities \cite{Cardy:2010zs,Ohmori:2014eia,Hellerman:2021fla}, or a description of the initial state in the global quantum quench \cite{Calabrese:2006rx}.

Another interesting aspect of BCFTs is their holographic dual.
According to the AdS/BCFT correspondence proposed in \cite{Takayanagi:2011zk, Fujita:2011fp, Karch:2000gx}, the boundary condition in BCFT is dual to what is called the End-of-the-World (ETW) brane, on which the bulk fields satisfy the Neumann boundary conditions, terminating the bulk space-time.\footnote{There are also explicit stringy constructions of AdS/BCFT with boundary conditions \cite{Gaiotto:2008sa,Aharony:2011yc,Chiodaroli:2011nr,Chiodaroli:2011fn,Chiodaroli:2012vc,Reeves:2021sab}.
}
ETW branes are also useful in analyzing the quantum nature of Black Holes, as some of the Black Hole microstates can be constructed from them \cite{Hartman:2013qma}.
The correspondence between Black Hole microstates and the boundary conditions (and hence their gravity dual, the ETW brane) has been useful in devising a toy model for the unitary Black Hole evaporation, qualitatively reproducing the Page curve obtained from the island formula in the gravity theory \cite{Almheiri:2019hni, Chen:2019uhq, Almheiri:2019psy, Balasubramanian:2020hfs, Geng:2020qvw, Chen:2020uac, Chen:2020hmv, Chen:2020jvn, Akal:2020twv, Miyaji:2021lcq, Akal:2021foz, Suzuki:2022yru,Suzuki:2022xwv, Izumi:2022opi}.

Motivated by the fact that the boundary conditions could partially correspond to Black Hole microstates,
a natural question would be how much we can distinguish one boundary condition from the other.
In particular, one wishes to find a physical quantity which distinguishes between boundary conditions.
Particularly in 2D CFTs, where there is a modular invariance, a natural candidate for such a quantity is the information metric defined on the space of boundary states.
The information metric
in general characterizes the distance between normalizable states, which expresses the probability of mistaking one state for the other.
Motivated by these, in this paper, we evaluate the information metric on the space of boundary states for various 2D CFTs.

Let us review the definition of the information metric briefly. (More details will be given in section~\ref{sec:definitions}.)
The quantum information theoretic distance between two normalisable pure states $d(\ket{\phi_1},\ket{\phi_2})$ are usually given as
\begin{align}
    d(\ket{\phi_1},\ket{\phi_2})=1-\abs{\braket{\phi_1|\phi_2}}.
\end{align}
The information metric $G_{ij}$ (usually called the Bures metric \cite{10.2307/1995012, Braunstein:1994zz}) is then defined by 
\begin{align}
    d(\ket{\lambda_k-\delta\lambda_k/2},\ket{\lambda_k+\delta\lambda_k/2})=G_{ij}\delta\lambda_i\delta\lambda_j+O(\delta\lambda^4),
\end{align}
for a family of parameterised states $\ket{\lambda_i}$.
As the boundary state is not normalisable in general, we will regularize it using the heat kernel 
\begin{align}
    \ket{B}\mapsto \ket{B}_\ell\equiv \frac{e^{-\ell H/2}\ket{B}}{\sqrt{\braket{B|e^{-\ell H}|B}}} \, ,
    \label{eq:regularization}
\end{align}
so that we can use the above definition of the information metric.
Here $\ket{B}$ is a boundary state and $\ket{B}_\ell$ is the regularized version with unit norm, where one usually takes $\ell\ll 1$.
The information metric computed from these states will tell us how different one boundary state compared to the other, in the quantum information theoretic sense.
Note that the boundary state $\ket{B}$ does not have to be conformal for the above definition to work, and we will study such a case in section~\ref{sec:nonconformal} as well.

There are several physical interpretations of the information metric of boundary states besides the quantum information theoretic one.
Imagine a conformal boundary condition preserving a continuous symmetry group $H$, which is a subgroup of the larger symmetry in the bulk, $G$.
It is then immediate that there is a family of conformal boundary states on the coset $G/H$.
In other words, the coset is nothing but the boundary conformal manifold.
Let us parametrize the coset by $\lambda$, and denote the corresponding boundary state as $\ket{B(\lambda)}$.
First of all, as we will check in the main body of the text, it is expected that the information metric on $G/H$ is proportional to the natural metric on the coset $G/H$ for symmetry reasons.
In the context of string theory, if we have a target space description of the 2D worldsheet CFT at hand, the information metric on the space of D-branes will probe the target space metric.

Second, this metric is nothing but the $O(\lambda^2)$ piece of the cylinder partition function $\braket{B(\lambda-\delta\lambda/2)|e^{-\ell H}|B(\lambda+\delta\lambda/2)}$ modulo multiplications by a constant.
(Note that the denominator of \eqref{eq:regularization} is independent of $\lambda$ for this specific case where the boundary conformal manifold is a coset.)
Now, by virtue of modular invariance,
\begin{align}
    \braket{B(\lambda_1)|e^{-\ell H}|B(\lambda_2)}=\Tr(e^{-\pi H_{\rm open}/\ell})
    = e^{-\pi E_0(\lambda_1,\lambda_2)/\ell}+O(e^{-E_1/\ell})
\end{align}
where $H_{\rm open}$ is the open string Hamiltonian defined on an interval with boundary conditions on the two ends and $E_0(\lambda_1,\lambda_2)$ ($E_1(\lambda_1,\lambda_2)$) is the system's ground state (first excited state) energy.\footnote{When the theory is fermionic, one will have to insert $(-)^F$ to the closed string amplitude. See \cite{Han:2017hdv,Smith:2019jnh,Smith:2020rru,Smith:2020nuf}. For general fermionic CFTs and their boundary conditions see \cite{Hsieh:2020uwb,Kulp:2020iet,Novak:2015ela,Runkel:2020zgg,Runkel:2022fzi,Smith:2021luc,Fukusumi:2021zme,Ebisu:2021acm,Burbano:2021loy}.}
$E_0(\lambda_1,\lambda_2)$ is obviously symmetric, so the $O(\delta\lambda^2)$ piece of $E_0(\lambda-\delta\lambda/2,\lambda+\delta\lambda/2)$ gives a change in the Casimir energy when we change the boundary conditions on an interval slightly.
To summarize the information metric on $G/H$ reflects the change in the Casimir energy on an interval when $\ell\ll 1$.

We can also extend the above to non-conformal boundary conditions.
Assuming non-conformal boundary states corresponding to such boundary conditions can be defined, as in \cite{Witten:1992cr,Bardakci:2001ck}, we can naturally extend the definition of the boundary information metric to non-conformal boundary states (but of CFTs, for simplicity).
In the main body of the text, we compute the information metric on the space of a boundary mass perturbation in free scalar theory in two dimensions.
This could potentially be interesting in the construction of the $g$-function governing the boundary RG flow.
The irreversibility of RG flows triggered by boundary relevant perturbations is an interesting issue, and the existence of the $g$-function was conjectured in \cite{Affleck:1991tk} and later proven in \cite{Friedan:2003yc,Casini:2016fgb} (See also \cite{ Dorey:1999cj,Yamaguchi:2002pa,Friedan:2003yc,Azeyanagi:2007qj,Takayanagi:2011zk,Estes:2014hka,Gaiotto:2014gha,Jensen:2015swa,Casini:2016fgb,Andrei:2018die,Kobayashi:2018lil,Casini:2018nym,Giombi:2020rmc,Wang:2020xkc,Nishioka:2021uef,Wang:2021mdq,Sato:2021eqo} for subsequent developments).
First of all, when there is only one relevant boundary parameter in the system, it is immediate to construct the restricted version of the $g$-function by integrating the metric times the beta function.
More interestingly, in section \ref{sec:nonconformal}, using the example above we numerically find that the boundary information metric computed (which is a scalar as we only considered only one relevant direction) is monotonic along the RG flow.
It would be interesting to study this further, but we leave it for future work.


Another interesting aspect of the boundary information metric is its holographic dual.
In \cite{Nozaki:2012zj,Miyaji:2015woj} the gravity dual of the information metric on the space of ordinary states was proposed.
They argued that it corresponds to the volume of the maximal time slice of the AdS spacetime under crude approximations.
Noting that the gravity dual of the boundary condition is the ETW brane \cite{Takayanagi:2011zk, Fujita:2011fp, Karch:2000gx}, a similar argument would result in the conclusion that it corresponds to the area of the extremal slice of the ETW brane connecting two boundaries of the asymptotic BCFT. In our case we should consider the minimal one not the maximal one because it is the point where the physically interesting things happen.
The only problem of this proposal is, however, that such a configuration for the ETW brane is absent when the two boundary conditions are different \cite{Miyaji:2021ktr}, which is exactly our case.
This conclusion is too naive.
The underlying assumption of \cite{Miyaji:2021ktr}, noticed recently, is that there is no degrees of freedom on the ETW brane \cite{Miyaji:2020dct,Biswas:2022xfw}.
This can be understood as follows:
Imagine a boundary state $\ket{B}$ of a theory with a $U(1)$ charge $Q$, and assume that $\ket{B}\neq e^{i\theta Q}\ket{B}$.
The proposal in \cite{Miyaji:2021ktr} indicates that the cylinder partition function discontinuously jumps at $\theta=0$, computed from its gravity dual.
This could be amended by realising that the $U(1)$ global symmetry on the CFT becomes the $U(1)$ gauge symmetry in the bulk dual, and that it effectively becomes a scalar field on the ETW brane.
Following this, we place a Klein-Gordon scalar field localized on the ETW brane, and argue that the action of the scalar field is the gravity dual of the boundary information metric.
We furthermore solve the EOM on the ETW brane and show that the profile of the scalar field is kink like,  therefore claiming that indeed the boundary information metric is the area of the minimal slice of the ETW brane connecting two boundaries of the asymptotic BCFT in a certain limit explained in the main body of the text.


The rest of the paper is organized as follows. 
We start in section \ref{sec:definitions} by giving the definition of distance and the metric in the space of (in particular boundary) states in QFTs. 

In section \ref{sec3} we show some examples of the calculation of the boundary information metric.
In section \ref{free} we start with a real free scalar theory with Neumann and Dirichlet boundary conditions in two-dimensional CFT and 
in section \ref{sec:wzw model} we compute the same quantity for spin 0 boundary states of $SU(2)_1$ Wess-Zumino-Witten (WZW) model and extend the result to the arbitrary level $k$ and spin $I$ case. We can recover the metric of the group manifold $SU(2)$.

In section \ref{sec:nonconformal}, we consider nonconformal boundary conditions of a CFT.
Concretely, we compute the information metric on the space of boundary mass perturbations in free scalar theory.

In section \ref{sec5} we conjecture, with supporting evidence, that the gravity dual of the boundary information metric is given by the action of the ETW with a scalar field localized on top.

In Appedix \ref{sec:d-branes} we rephrase the result in Sec. \ref{sec3} in terms of string theory and D-branes. We also comment on the behavior of the information metric in the long cylinder limit.

\section{Definitions}
\label{sec:definitions}
We define the notion of distance and metric on the state space of Quantum Field Theories (QFTs).
Let's say we are interested in measuring how close to each other two (normalised) density matrices, $\r$ and $\r'$, are.
One measure of such a quantity is called the fidelity (For reviews and the information theoretic meaning of fidelity see \cite{doi:10.1080/09500349414552171}.) and is defined as follows,
\be
F=\operatorname{Tr}\left( \sqrt{\sqrt{\rho}\rho'\sqrt{\rho}} \right).
\ee
Similarly, the affinity $A$ and the trace norm $I$ can be defined as in \cite{PhysRevA.69.032106,WANG200858} by
\begin{align}
    A&=\operatorname{Tr}[\sqrt{\rho}\sqrt{\rho'}] \, , \\[2pt]
    I&=\frac{\operatorname{Tr}\rho\rho'}{\sqrt{(\operatorname{Tr}\rho^2)(\operatorname{Tr}\rho^{'2})}} \, .
\end{align}
For pure states, we can write the density matrices as $\rho=\ket{\phi}\bra{\phi}$ and $\rho'=\ket{\phi'}\bra{\phi'}$, and the above quantities become
\begin{align}
    F \, = \, A \, = \, \abs{\braket{\phi|\phi'}} \, , \qquad 
    I \, = \, \abs{\braket{\phi|\phi'}}^2 \, .
\end{align}
The distance corresponding to affinity and trace norm is called the Bures and the Hellinger distance,
\begin{align}
    D_B^2(\rho(\lambda+\delta\lambda/2),\rho(\lambda-\delta\lambda/2))&=2(1-F)=G^B_{ij}d\lambda^i d\lambda^j+\cdots,\nn\\
     D_H^2(\rho(\lambda+\delta\lambda/2),\rho(\lambda-\delta\lambda/2))&=2(1-A)=G^H_{ij}d\lambda^i d\lambda^j+\cdots.
\end{align}
We will call  $G^B_{\lambda\lambda}$ and $G^H_{\lambda\lambda}$ the Bures  and the Hellinger metric, respectively.
Note that this expansion series in $\delta\lambda$ starts at the quadratic order.

Now, we consider the information metric on the space of boundary states. 
We will hereafter only concern ourselves with the pure boundary states, although it is perfectly possible to consider mixed boundary states.\footnote{Mixed boundary states have recently been considered in holographic contexts in \cite{Suzuki:2019xdq,Kusuki:2019hcg}.}
The obstacle in defining the boundary information metric is that the boundary states are never normalizable, but we can normalise it by using the heat kernel regularization,
\begin{align}
    \ket{B}\mapsto \ket{B}_\ell\equiv \frac{q^{-H/2}\ket{B}}{\sqrt{\braket{B|q^{-H}|B}}} \, ,
    \label{eq:regularizationq}
\end{align}
which is nothing but \eqref{eq:regularization} with $q\equiv e^{\ell}$.
Writing the cylinder partition function with boundary conditions $B(\lambda)$ and $B(\lambda')$ as
\be
Z_{\lambda\lambda}=\bra{B(\lambda)}q^{-H_{closed}}\Ket{B(\lambda)} \, ,
\ee
the Bures and the Hellinger distance (and consequently metric) can be defined as
\begin{align}
    D_B^2=D_H^2=2-2\abs{\frac{Z_{\lambda''\lambda'}}{\sqrt{Z_{\lambda''\lambda''}}\sqrt{Z_{\lambda'\lambda'}}}}_{\lambda'=\lambda+d\lambda,\lambda''=\lambda-d\lambda}= G^B_{ij}d\lambda^i d\lambda^j+\cdots,\label{37}
\end{align}
where we take the second order in the expansion series. 
Note that the Bures and the Hellinger metric coincides with each other when we consider pure states.

\section{Examples}\label{sec3}
\subsection{Free Scalar CFT$_2$} \label{free}
We study two-dimensional free scalar theory defined by
    \begin{align}
        I \, = \, \frac{1}{2} \int_0^\inf dw \int_{-\inf}^\inf d\t \Big[ (\pa_\t X )^2 + (\pa_w X )^2 \Big] \, .
    \label{I_bulk}
    \end{align}
We have a straight-line boundary at $w=0$ and we use string theory notation as follows \cite{Callan:1987px,Blumenhagen:2009zz}.
The variation of the action leads to 
    \begin{align}
        \d I \, = \, - \int_0^\inf dw \int_{-\inf}^\inf d\t \, \d X ( \pa_\t^2 + \pa_w^2 ) X \, + \, \int_{w=0} d\t \, \d X \pa_w X \, .
    \end{align}
Therefore, to make the variation principle well defined, the allowed boundary conditions at $w=0$ are 
    \begin{align}
        \textrm{Neumann}:& \qquad \pa_w X \big|_{w=0} \, = \, 0 \, , \\
        \textrm{Dirichlet}:& \qquad \d X\big|_{w=0} \, = \, \pa_\t X \big|_{w=0} \, = \, 0 \, .
    \end{align}

Since we have $U(1)$ currents for this theory:
    \begin{align}
        j(z) \, = \, i \, \pa X \, = \, \sum_{n \in \mathbb{Z}} j_n \, e^{-nz} \, , \qquad
        \bar{j}(\bar{z}) \, = \, i \, \bar{\pa} X \, = \, \sum_{n \in \mathbb{Z}} \bar{j}_n \, e^{- n\bar{z}} \, ,
    \end{align}
it's convenient to use the current algebra:
    \begin{align}
        \big[ j_m, j_n \big] \, = \, m \, \d_{m, -n} \, ,
    \label{current_alg}
    \end{align}
where we defined $z=w+i \t$ and $\pa = \frac{1}{2}(\pa_w - i \pa_\t)$.
Therefore, we have
    \begin{align}
        i \, \pa_w X \big|_{w=0} \, &= \, \sum_{n\in \mathbb{Z}} \Big( j_n e^{-in\t} + \bar{j}_n e^{in \t} \Big) \, , \\
        \pa_\t X \big|_{w=0} \, &= \, \sum_{n\in \mathbb{Z}} \Big( j_n e^{-in\t} - \bar{j}_n e^{in \t} \Big) \, .
    \end{align}
The corresponding boundary states $| B \ra$ are defined by
    \begin{align}
        \textrm{Neumann}:& \qquad (j_n + \bar{j}_{-n}) | B_{\textrm{N}} \ra \, = \, 0 \, , \\
        \textrm{Dirichlet}:& \qquad (j_n - \bar{j}_{-n}) | B_{\textrm{D}} \ra \, = \, 0 \, .
    \end{align}
These conditions are solved by the coherent states \cite{Callan:1987px} as
\footnote{Boundary states in free fermion theory was obtained in \cite{Polchinski:1987tu}. For a review, see \cite{DiVecchia:1999mal}.}
    \begin{align}
        \textrm{Neumann}:& \qquad | B_{\textrm{N}} \ra \, = \, \frac{1}{\mathcal{N}_{\textrm{N}}} \,
        \exp\left( - \sum_{k=1}^\inf \frac{1}{k} \, j_{-k} \bar{j}_{-k} \right) | 0 \ra \, , \label{B_N} \\
        \textrm{Dirichlet}:& \qquad  | B_{\textrm{D}} \ra \, = \, \frac{1}{\mathcal{N}_{\textrm{D}}} \,
        \exp\left( + \sum_{k=1}^\inf \frac{1}{k} \, j_{-k} \bar{j}_{-k} \right) | 0 \ra \, , \label{B_D}
    \end{align}
where $\mathcal{N}_{\textrm{N}}= \sqrt{2}$ and $\mathcal{N}_{\textrm{D}}=1$.

By using 
    \begin{align}
        L_n \, &= \, \frac{1}{2} \sum_{k>-1} j_{n-k} \, j_k \, + \, \frac{1}{2} \sum_{k\le -1} j_k \, j_{n-k} \, , \\
        \overline{L}_n \, &= \, \frac{1}{2} \sum_{k>-1} \bar{j}_{n-k} \, \bar{j}_k \, + \, \frac{1}{2} \sum_{k\le -1} \bar{j}_k \, \bar{j}_{n-k} \, ,
    \end{align}
and also noting that 
    \begin{align}
        j_0 | B_{\textrm{N,D}} \ra \, = \, \bar{j}_0 | B_{\textrm{N,D}} \ra \, = \, 0 \, ,
    \end{align}
we can explicitly check that the Cardy condition is satisfied 
    \begin{align}
        \big( L_n - \overline{L}_{-n} \big) | B_{\textrm{N,D}} \ra \, = \, 0 \, .
    \end{align}

The Neumann-Neumann cylinder amplitude with cylinder length $l$ is given by 
    \begin{align}
        Z_{(N,N)}(l) \, = \, \frac{1}{\mathcal{N}_N^2} \, \frac{1}{\h(2il)} \, ,
    \end{align}
while the Dirichlet-Dirichlet cylinder amplitude is given by 
    \begin{align}
        Z_{(D,D)}(l) \, = \, \frac{1}{\sqrt{2l}\mathcal{N}_D^2} \, \frac{1}{\h(2il)} \, \exp\left( - \frac{(x_a - x_b)^2}{8\pi l} \right) \, ,
    \end{align}
where $x_a$ and $x_b$ are the boundary values of the scalar field $X$. As usual, we defined $\h(2il)=q^{\frac{1}{24}}\prod_{n=1}^\inf(1-q^n)$ with $q=e^{-4\pi l}$.

In the $x_a \to x_b$ limit, regarding $\d\l = x_a - x_b$, we find
    \begin{align}
        D_B^2 \, = \, \frac{\d\l^2}{4\pi l}+O(\delta\lambda^4) \, .
    \end{align}
As we will see in section~\ref{sec:d-branes}, if we regard this theory as a worldsheet theory of an open string stretched between two D-branes,
this metric coincides with the flat target space metric on the D-brane.

\subsection{SU(2) WZW Model}
\label{sec:wzw model}
As a second example, in this section we demonstrate our idea that the boundary information metric probes the metric of the target space.
Here, we consider SU(2)-level 1 Wess-Zumino-Witten (WZW) model and show that the boundary information metric matches with that of SU(2). We firstly restrict to the spin zero case and generalize our discussion to the arbitrary level $k$ and spin $I$ case at the end of this section.
 
The action of the WZW model on Riemann surface $\Sigma$ is given by 
\begin{align}
    I=\frac{k}{8\pi}\int_\Sigma d^2x \mathrm{Tr}(\partial_\alpha g\partial_\alpha g^{-1})-\frac{ik}{12\pi}\int_B \mathrm{Tr}(g^{-1} dg\wedge g^{-1} dg\wedge g^{-1} dg),
\end{align}
where $g$ is a map from $\Sigma$ to the Lie group $G$, and $B$ is an arbitrary 3-dimensional manifold with boundary $\Sigma$.
This action is well-defined, independent to the choice of $B$ if $k$ is integer. $k$ is restricted to positive to make the action bounded below.
The WZW model is a 2-dimensional CFT, with the symmetry of current algebra.
The currents are giben by
\begin{align}
    J=&~J^aT_a=-k\partial g g^{-1} \, , \\
    \bar{J}=&~\bar{J}^aT_a=kg^{-1}\bar{\partial} g \, ,
\end{align}
where $T_a$ is the generator of $G$.
Taking the Laurent expansion of these currents
\begin{subequations}
\begin{align}
    J^a(z)=&~\sum_{n\in\mathbb{Z}}\frac{J^a_n}{z^{n+1}}      ,   
    \\
    \Bar{J}^a(\Bar{z})=&~\sum_{n\in\mathbb{Z}}\frac{\Bar{J}^a_n}{\Bar{z}^{n+1}}      ,  
\end{align}
\end{subequations}
their modes, $J_n^a$ and $\Bar{J}_n^a$, satisfy current algebras: 
\begin{subequations}
\begin{align}
    \left[J_n^a,J_m^b\right] \, &= \, if^{ab}{}_c J^c_{n+m} + k n \, \d^{ab} \delta_{n+m,0},\\
    \left[\Bar{J}_n^a,\Bar{J}_m^b\right] \, &= \, if^{ab}{}_c \Bar{J}^c_{n+m} + k n \, \d^{ab} \delta_{n+m,0}.
\end{align}
\end{subequations}
In our setup we set $G=SU(2)$ and we consider $k=1$ case for a while.\footnote{We consider the diagonal model (A-series in ADE classification).}

In the CFT with current algebra symmetry, boundary states must satisfy the gluing conditions
\begin{align}
    (L_n-\Bar{L}_{-n})|B\rangle=0, \\
    (J^a_n + \Omega\Bar{J}^a_{-n})|B\rangle=0,
\end{align}
where $\Omega$ is an automorphism of the algebra. In the case of SU(2) WZW model, we can rotate $\Bar{J}^a_{-n}$ by $g\in SU(2)$. The choice of $g$ is arbitrary, thus gives the moduli space of boundary states.
In \cite{Gaberdiel:2001xm}, a spin 0 boundary state in the level-$1$ SU(2) WZW model with gluing condition $g$ is given by
 \be
 \Ket{B,g}=\frac{1}{2^{\frac{1}{4}}}\sum_{(j,m,n)} D^j_{m,n}(g) |j,m,n\rangle\rangle,
 \ee
where $|j,m,n\rangle\rangle$ is Virasoro Ishibashi state, $m,n=-j,\cdots,j$ and $j$ labels eigenvalues of $J_z$. The explicit form of Klebsch-Gordan coefficients are
\be
 D^j_{m,n}(g)=\sum_{l=max(0,n-m)}^{min(j-m,j+n)}\frac{\sqrt{(j+m)!(j-m)!(j+n)!(j-n)!}}{(j-m-l)!(j+n-l)!l!(m-n+l)!}a^{j+n-l}d^{j-m-l}b^lc^{m-n+l},
\ee
where we represent $g\in SU(2)$ as
\be
g \, = \, 
\begin{pmatrix}
a & b\\
 c & d\\
\end{pmatrix}.
\ee
We originally consider a cylinder amplitude between $g_1$ and $g_2$ with circumference $R$ and length $L$. By a conformal mapping, it can be mapped to an annulus with an inner radius  $\exp(-2\pi L/R)$ and an outer radius 1.
The annulus partition function can be represented in terms of Klebsch-Gordan coefficients and characters 
\begin{align}
    \mathcal{A}=&\bra{B,g_1}q^{\frac{1}{2}(L_0+\bar{L_0}-\frac{c}{12})}  \Ket{B,g_2}\nn\\
    =&\frac{1}{\sqrt{2}}\sum_{j\in\frac{\mathbb{Z}}{2}}\sum_{m,n}(-1)^{-m,-n} D^j_{m,n}(g_1) D^j_{m,n}(g_2)\chi_{j^2}(q),
\end{align}
where 
\begin{align}
    q&=e^{-4\pi L/R}, \\ \chi_h(q)&=\theta_{\sqrt{2}j}(q)-\theta_{\sqrt{2}(j+1)}(q), \\
    \theta_s(q)&=\frac{q^{\frac{s^2}{2}}}{q^{\frac{1}{24}}\prod_{n=1}^{\infty}(1-q^n)}.
\end{align}

\begin{figure}
    \begin{center}
        \includegraphics[width=70mm]{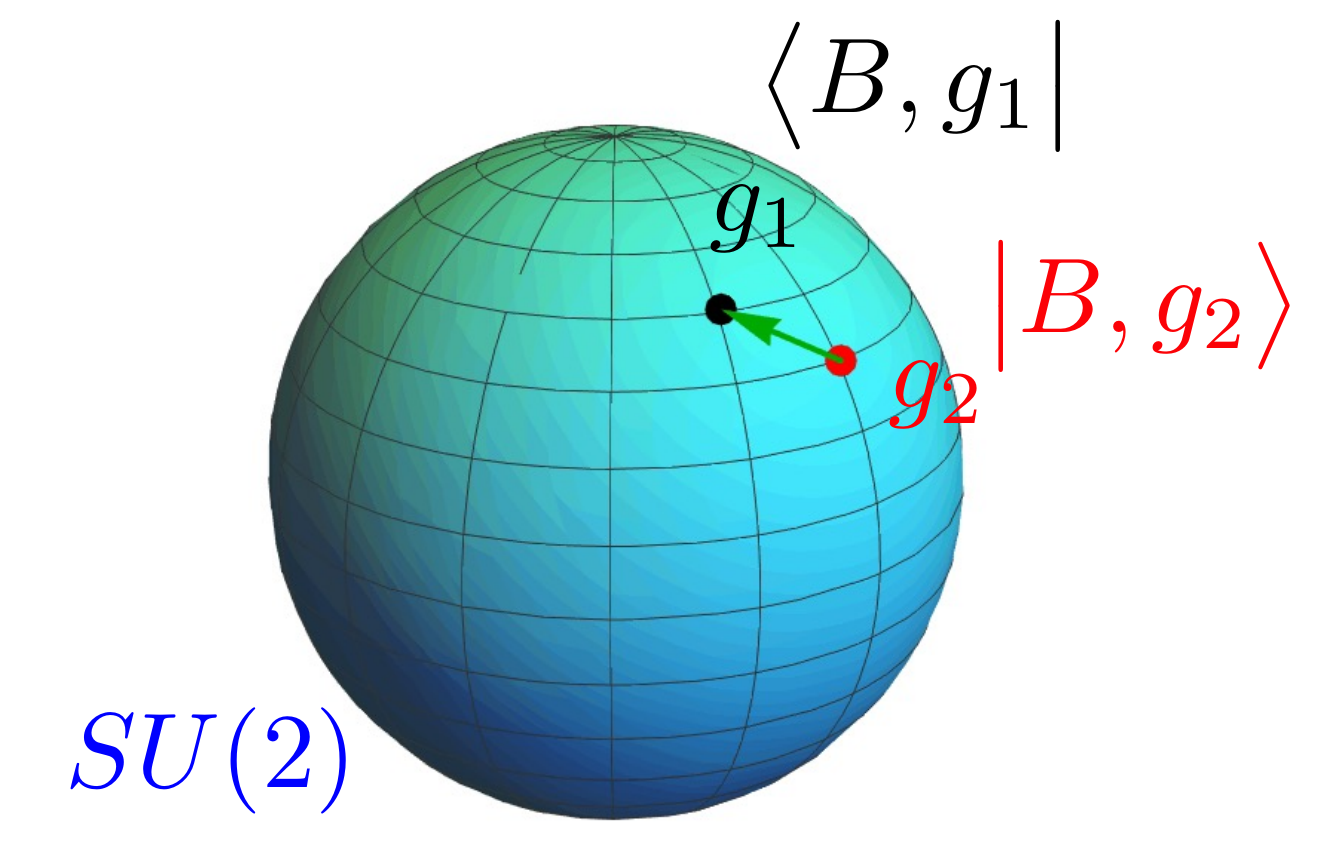}
        \vspace*{-10pt}
    \end{center}
    \caption{The amplitude between two boundary states on $SU(2)$.}
\end{figure}

The sum of the representation matrices gives
 \be
 \sum_{m,n} D^j_{m,n}(g_1^{-1})D^j_{m,n}(g_2)=\sum_n D^j_{n,n}(g_1^{-1}g_2).
 \ee
If we take the Jordan normal form of $\hat{g}=g_1^{-1}g_2$, then we obtain
 \be
 D^j_{n,n}(\hat{g})=\hat{a}^{2n},
 \ee
where $\hat{g}=\begin{pmatrix}
\hat{a} & 0\\
 0 & 1/\hat{a}\\
\end{pmatrix}\in SU(2)$. After a simple algebra we obtain
\be
\mathcal{A}=\sum_{n} \frac{\tilde{q}^{n^2+\frac{i n\alpha}{\pi}-\frac{\alpha^2}{4\pi^2}}}{\eta(\tilde{q})},
\ee
where $\hat{a}=e^\alpha$ and $\tilde{q}=e^{-\frac{\pi R}{L}}$.

Let us now calculate the information metric.  We should evaluate the  amplitude with the same position in the group manifold, where we set $\alpha=0$
\be
Z_{gg}=\sum_n \frac{\tilde{q}^{n^2}}{\eta(\tilde{q})}.
\ee
We can also calculate a perturbed amplitude for an infinitesimal $\alpha$
\be
Z_{gg'}=\sum_{n} \frac{\tilde{q}^{n^2+\frac{i n\alpha}{\pi}-\frac{\alpha^2}{4\pi^2}}}{\eta(\tilde{q})}.
\ee
Then, the Bures distance is given by
\be
\label{eq:su2level1BuresMetric}
D_B^2=2\left(1-\abs{\frac{Z_{gg'}}{Z_{gg}}}\right)=2(1-e^{\frac{R\alpha^2}{4\pi L}})\sim\frac{R\beta^2}{2\pi L},
\ee
where $\beta=i\alpha$.

To derive a metric,  we denote the $g$ and $g'$ as
$g=\begin{pmatrix}
e^{i\xi_1}\sin\eta & e^{i\xi_2}\cos\eta\\
 -e^{-i\xi_2}\cos\eta & e^{-i\xi_1}\sin \eta\\
\end{pmatrix}$
and
$g'=\begin{pmatrix}
e^{i\xi'_1}\sin\eta' & e^{i\xi'_2}\cos\eta'\\
 -e^{-i\xi'_2}\cos\eta' & e^{-i\xi'_1}\sin \eta'\\
\end{pmatrix}$, respectively. This is called Hopf coordinates of $SU(2)$.
After we diagonalize the matrix $g^{-1}g'$ we obtain 
\be
\beta\sim \sqrt{(\sin^2\eta (\Delta\xi_1^2)+\cos^2\eta (\Delta\xi_2)^2+(\Delta\eta)^2)},
\ee
where we define $\Delta\xi_1=\xi'_1-\xi_1$, $\Delta\xi_2=\xi'_2-\xi_2$ and $\Delta\eta=\eta'-\eta$. Here we also ignored higher-order terms in $\Delta\xi_i$.
From this the information metric becomes
\be
D_B^2=\frac{R}{2\pi L}(d\eta^2+\sin^2\eta d\xi_1^2+\cos^2\eta d\xi_2^2),
\ee
which reproduces the metric of SU(2) in Hopf coordinates. We note that in the closed string picture the modulus is $L/R$ and the boundary information metric is proportional to an inverse of the modulus.

As one can see, in the short cylinder limit $|Z_{gg'}/Z_{gg}|$ picks up $\tilde{q}^h$, where $h$ is a conformal weight of the open-string vacuum with boundary conditions $g$ and $g'$. This means that when we calculate the information metric in the short cylinder limit we do not need to explicitly construct the set of boundary states if we know $h$. In the case of SU(2)-level $k$ WZW model, $h$ corresponding to the spin 0 representation of $g$ and $g'$ is determined in \cite{Elitzur:2000pq}:
\begin{align}
\label{eq:generalh}
h = k\left( \frac{\beta}{2\pi} \right)^2.
\end{align}
The Bures distance is
\begin{align}
D^2_B\sim \frac{kR\beta^2}{2\pi L},
\end{align}
here $k=1$ corresponds to (\ref{eq:su2level1BuresMetric}). 
More generically, $Z_{gg'}$ for generic spin (labelled by $I,J$) representation is found in \cite{Recknagel:2013uja}:
\begin{align}
    Z_{(g,I)(g',J)}=\sum_j N^j_{IJ}\Tr_{\mathcal{H}_j}\Tilde{q}^{L_0-\frac{c}{24}-\frac{2\beta}{2\pi} J^3_0+ k\left(\frac{\beta}{2\pi}\right)^2}.
\end{align}
We can read off the open-string vacuum contribution $h$ for general spin $I=J$ to be (\ref{eq:generalh}).


\section{Boundary Mass Perturbations}
\label{sec:nonconformal}
So far in this paper, we have been considering BCFTs which keep maximal subgroup of the dulk conformal symmetry at its boundary.
Such maximal subgroup on the boundary is called boundary conformal symmetry.
For these cases, the parameter $\l$ is the coordinate of the boundary conformal manifold.
Now, we study the case when we break this boundary conformal symmetry and take $\l$ as the breaking parameter.
In particular in this section, we consider an example of such breaking by adding boundary mass perturbations.
In the context of string theory, such boundary mass perturbations were studied as the so called boundary string field theory \cite{Witten:1992cr, Li:1993za, Kutasov:2000qp, Bardakci:2001ck, Arutyunov:2001nz}.

\subsection{Bosonic Case}

In this subsection, we consider a cylinder amplitude of scalar fields with a modified Neumann condition. In the context of string field theory this was already studied in \cite{Witten:1992cr,Bardakci:2001ck}. In this section we will follow the formalism of \cite{Bardakci:2001ck}.

We consider a free scalar with a following boundary term
\be
S=-\frac{1}{4\pi\alpha'}\int_{0}^{2\pi} d\sigma\int_{-\frac{\pi}{t}}^0d\tau\partial_\mu X \partial^\mu X-\frac{\lambda\delta(\tau=0)}{4\pi\alpha'}\int_0^{2\pi}d\sigma X^2-\frac{\lambda\delta(\tau=-\frac{\pi}{t})}{4\pi\alpha'}\int_0^{2\pi}d\sigma X^2.
\ee
The boundary condition at $\tau=0,-\pi/t$ becomes
\be
(\partial_{\tau'} X+t\lambda X)|_{\tau=0,-\frac{\pi}{t}}=0,\label{629}
\ee
which we call a modified Neumann condition.

 The mode expansion in complexified coordinates $z=e^{\tau+i\sigma}$ becomes 
\be
X=X_0-\frac{ip\alpha'}{2}\log\abs{z}^2+i\sqrt{\frac{\alpha'}{2}}\sum_m \frac{1}{\sqrt{m}}\left(\alpha_m e^{-m(\tau+i\sigma)}+\tilde{\alpha}_m e^{-m(\tau-i\sigma)}\right).\label{630}
\ee
The commutation relation is given by
\begin{align}
    [X_0,p]&=i\nn\\
    [\alpha_m,\alpha_{-n}]&=\delta_{mn}.
\end{align}
Plugging (\ref{630}) into (\ref{629}), we obtain the boundary conditions on the left and right creation and annihilation operators
\be
\left.\left(\alpha_m-\frac{m-t\lambda}{m+t\lambda}e^{2m\tau}\tilde{\alpha}_{-m}\right)\right|_{\tau=-\frac{\pi}{t},0}=0.
\ee
Foe general $\tau$ the boundary states can be described by
\begin{align}
    \alpha_m\Ket{\lambda}_\tau&=\frac{m-t\lambda}{m+t\lambda}e^{2m\tau}\tilde{\alpha}_{-m}\Ket{\lambda}_\tau\nn\\
    \tilde{\alpha}_m\Ket{\lambda}_\tau&=\frac{m-t\lambda}{m+t\lambda}e^{2m\tau}\alpha_{-m}\Ket{\lambda}_\tau\nn\\
    i\hat{p}\Ket{\lambda}_\tau&=-\frac{t\lambda\hat{X}_0}{\alpha'(1-t\lambda\tau)}\Ket{\lambda}_\tau,
\end{align}
where $\Ket{\lambda}_\tau$ represents the boundary state at general $\tau$.
We can find the boundary state, which satisfies the above conditions
 \be
 \Ket{\lambda}_\tau=\mathcal{N}_{\lambda}\sqrt{\frac{2t\lambda}{\alpha'(1-t\lambda\tau)}}\exp\left(\sum_{m=1}^{\infty}(\frac{m-t\lambda}{m+t\lambda})\alpha_{-m}\tilde{\alpha}_{-m}e^{2m\tau}\right)\exp{\left(-\frac{t\lambda \hat{X_0^2}}{2\alpha'(1-t\tau\lambda)}\right)}\Ket{0}.
 \ee
 We can fix the normalization as follows. The tree-level partition function factor, which corresponds to the disk amplitude, was found in \cite{Witten:1992cr} to be 
 \be
 Z_0=\left(\sqrt{t\lambda}e^{\gamma t\lambda}\Gamma(t\lambda)\right)^{\frac{2}{\al}}.
 \label{Z0}
 \ee
 Then, we have 
 \be
 \mathcal{N}_{\lambda}=N_0 Z_0,
 \ee
 where $N_0$  can be fixed by comparing to the open string channel and $N_0=\frac{1}{\sqrt{2}}$.
 
 The cylinder amplitude between boundary states at $\tau=-\pi/t$ and $\tau=0$ becomes
\begin{align}
Z&=\bra{\lambda}_{\tau=0}\Ket{\lambda}_{\tau=-\pi/t}\nn\\
&=N_0^2 Z_0^2 \sqrt{\frac{2t\lambda}{\pi\al(2+\pi\lambda)}}\prod_{m=1}^{\infty}\frac{1}{1-\left(\frac{m-t\lambda}{m+t\lambda}\right)^2 e^{-\frac{2\pi m}{t}}}\nn\\
&=\frac{1}{2}\left(\sqrt{t\lambda}\Gamma(t\lambda)e^{t\lambda\gamma}\right)^{\frac{4}{\alpha'}}\sqrt{\frac{2t\lambda}{\pi\al(2+\pi\lambda)}}\prod_{m=1}^{\infty}\frac{1}{1-\left(\frac{m-t\lambda}{m+t\lambda}\right)^2 e^{-\frac{2\pi m}{t}}},
\label{cylinder_amp}
\end{align}
where $\gamma$ is the Euler's constant.
We note that in the $\l \to 0$ limit, $Z$ diverges as $Z \propto \l^{-1/2}$.
This is because the Neumann boundary condition (i.e. strictly at $\l=0$) is distinct from the $\l \to 0$ limit.
For the Neumann boundary condition, the boundary state is independent of the zero mode $\hat{X}_0$, so the corresponding amplitude excludes the zero mode integral.
On the other hand, for the modified Neumann case ($\l >0$) we have the zero mode integral
    \begin{align}
        \int_{-\inf}^\inf \frac{d\hat{X}_0}{2\pi} \, \exp\left(- \frac{t \l}{2\a'} \hat{X}_0^2\right) \exp\left(- \frac{t \l}{2\a'(1-t \l)} \hat{X}_0^2\right)
        \, = \, \sqrt{\frac{\a'(1-t \l)}{2\pi t \l (2-t\l)}} \, ,
    \end{align}
in the amplitude (\ref{cylinder_amp}). This Gaussian integral is responsible for the $Z \propto \l^{-1/2}$ divergence.

\begin{figure}
    \centering
        \includegraphics[width=8cm]{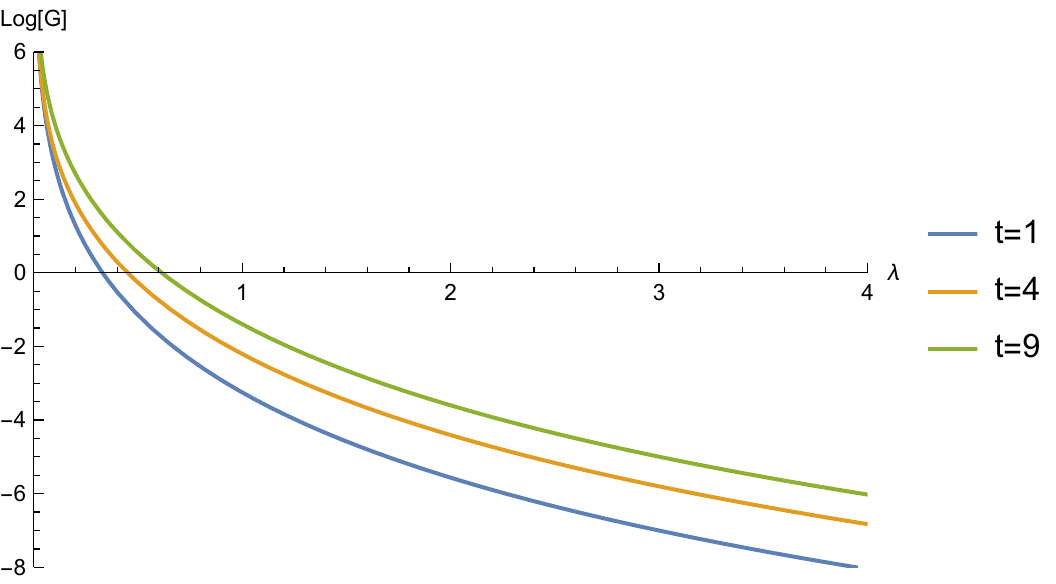}
        \caption{Log plot of the information metric $G$ with $t=1,4,9$, where we chose $\a'=2$ and $m_{\textrm{max}}=10$.
        As we explain around table~\ref{table:m}, $m_{\textrm{max}}=10$ is sufficient for the case when $t=\mathcal{O}(1)$ and $\l=\mathcal{O}(1)$.
        For larger $t$, higher $m_{\textrm{max}}$ we need.}
\label{fig:infometric}
\end{figure}

For the calculation of the information metric, we calculate the amplitude from the boundary condition $\lambda$ to $\lambda'$, which is slightly perturbed from $\lambda$. The partition function becomes
\begin{align}
    Z&=\bra{\lambda'}_{\tau=0}\Ket{\lambda}_{\tau=-\pi/t}\nn\\
    &=\frac{1}{2}\!\left(\sqrt{t\lambda}\Gamma(t\lambda)e^{t\lambda\gamma}\right)^{\frac{2}{\alpha'}}\!\left(\sqrt{t\lambda'}\Gamma(t\lambda')e^{t\lambda'\gamma}\right)^{\frac{2}{\alpha'}}\!\sqrt{\frac{2t\lambda^2}{\al\pi(\lambda+\lambda'+\lambda\lambda'\pi)}}\nn\\
    &\cdot\prod_{m=1}^{\infty}\frac{1}{1-\left(\frac{m-t\lambda}{m+t\lambda}\right)\left(\frac{m-t\lambda'}{m+t\lambda'}\right) e^{-\frac{2\pi m}{t}}}.
\label{z_amplitude}
\end{align}
From this amplitude, the information metric is then computed by
    \begin{align}
        G(\l) \, = \, \lim_{\epsilon\to 0}\frac{1-P(\l+\e, \l-\e)}{\e^2} \, , \qquad 
        P(\l_1, \l_2) \, = \, \frac{Z(\l_1, \l_2)}{\sqrt{Z(\l_1, \l_1) Z(\l_2, \l_2)}} \, .
    \end{align}
We can obtain an analytical expression, but the result is lengthy, so here we instead show numerical plots in Figure~\ref{fig:infometric},
as well as the disk amplitude behavior.
These plots show that the information metric monotonically decreases as $\l$ grows. 

For the numerical evaluation of the information metric, we need to put an upper cutoff $m_{\textrm{max}}$ on the infinite product in the amplitude
    \begin{align}
        Z(\l_1, \l_2) \, = \, A(\l_1, \l_2) \prod_{m=1}^{m_{\textrm{max}}} B_m(\l_1, \l_2) \, ,
    \end{align}
where
    \begin{align}
        A(\l_1, \l_2) \, &= \, \frac{1}{2}\!\left(\sqrt{t\l_1}\Gamma(t\l_1)e^{t\l_1\gamma}\right)^{\frac{2}{\alpha'}}\!\left(\sqrt{t\l_2}\Gamma(t\l_2)e^{t\l_2\gamma}\right)^{\frac{2}{\alpha'}}\!\sqrt{\frac{2t\l_1 \l_2}{\al\pi(\l_1+\l_2+\l_1\l_2\pi)}} \, , \nn\\
        B_m(\l_1, \l_2) \, &= \, \left[ 1-\left(\frac{m-t\lambda}{m+t\lambda}\right)\left(\frac{m-t\lambda'}{m+t\lambda'}\right) e^{-\frac{2\pi m}{t}} \right]^{-1} \, .
    \end{align}
In table~\ref{table:m}, we show that $B_m(\l_1, \l_2)$ is very quickly suppressed as $m$ grows, at least for $t=\mathcal{O}(1)$ and $\l=\mathcal{O}(1)$.

\begin{table}[t!]
    \centering
    \begin{tabular}{|c||c|c|c|c|} 
        \hline
        $m$ & $ B_m|^{t=2.2}_{\l_1=\l_2=2.3} -1$ & $ B_m|^{t=4.6}_{\l_1=\l_2=2.3}-1$ & 
        $ B_m|^{t=2.2}_{\l_1=\l_2=8.7}-1$ & $ B_m|^{t=4.6}_{\l_1=\l_2=8.7}-1$ \\ [0.5ex] \hline
        1 & 2.649 $\times$ $10^{-2}$ & 2.116 $\times$ $10^{-1}$ & 4.893 $\times$ $10^{-2}$ & 3.002 $\times$ $10^{-1}$ \\ 
        2 & 6.215 $\times$ $10^{-4}$ & 3.123 $\times$ $10^{-2}$ & 2.178 $\times$ $10^{-3}$ & 5.630 $\times$ $10^{-2}$ \\
        3 & 1.242 $\times$ $10^{-5}$ & 5.202 $\times$ $10^{-3}$ & 1.010 $\times$ $10^{-4}$ & 1.245 $\times$ $10^{-2}$ \\
        4 & 1.496 $\times$ $10^{-7}$ & 8.639 $\times$ $10^{-4}$ & 4.679 $\times$ $10^{-6}$ & 2.846 $\times$ $10^{-3}$ \\
        \hline
    \end{tabular}
    \caption{Suppression in $m$ for $B_m-1$. We can see for $t=\mathcal{O}(1)$ and $\l=\mathcal{O}(1)$, the suppression is very fast.}
    \label{table:m}
\end{table}

\begin{figure}
    \centering
        \includegraphics[width=7.5cm]{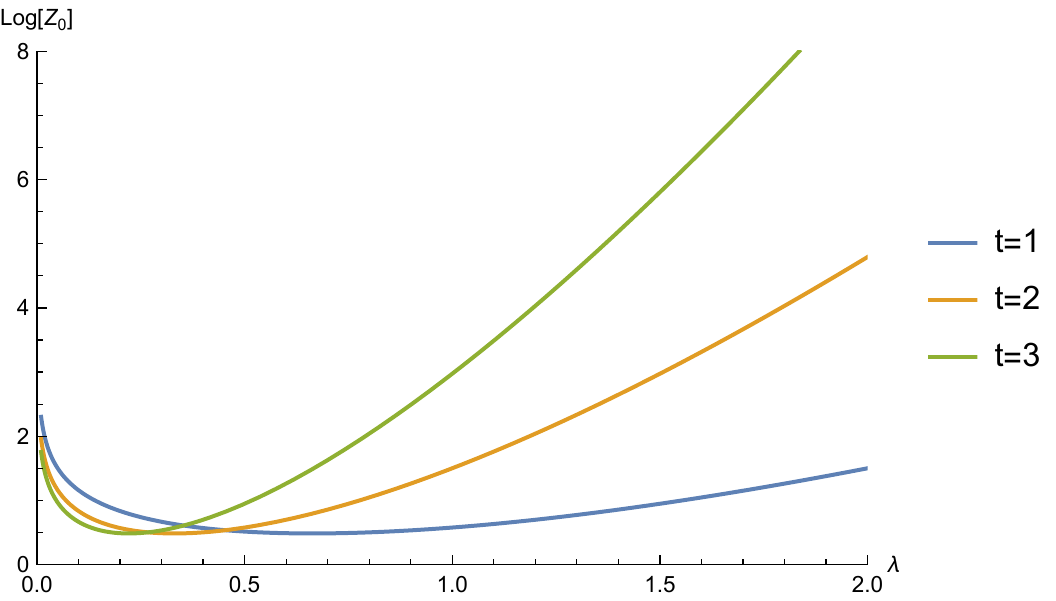} \ \ \includegraphics[width=7.5cm]{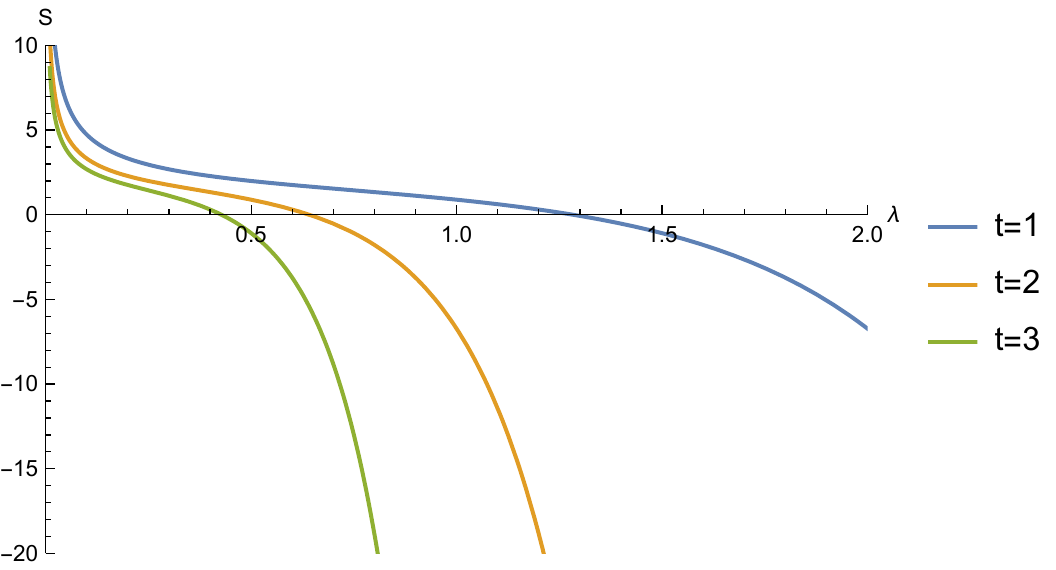}
        \caption{Left: Log plot of the disk amplitude $Z_0$ with $t=1,2,3$.
        Right: Log plot of the  action $S$ defined in (\ref{S}) for $t=1,2,3$.}
\label{fig:Z0&S}
\end{figure}

The result that the information metric (which is a scalar in our case) is monotonic along the RG flow in our case reminds us of the $c$-function \cite{Zamolodchikov:1986gt}.
The boundary version of such an object is given by
the action $S$ discussed in \cite{Kutasov:2000qp,Affleck:1991tk}. This action is related to the disk partition function (\ref{Z0}) by
    \begin{align}
        S \, = \, \left( 1- \l \frac{\pa}{\pa \l}\right) Z_0 \, .
    \label{S}
    \end{align}
We plot $S$ and $Z_0$ for our case in figure~\ref{fig:Z0&S}.
The apparent monotonicity in our example seems to indicate that it is easier to distinguish boundary conditions with larger degrees of freedom.
Unfortunately though, we could not find any concrete relationship between the two.
Nevertheless, we have an upper bound of the information metric which monotonically decreases along the RG trajectories \cite{Casini:2018cxg}.
This does not prove the information metric itself monotonically decreases
\footnote{In particular, the information metric obtained from the quantum Renyi relative entropies in \cite{Casini:2018cxg} is defined by ordinary states,
while we are discussing the information metric defined from boundary states. We will discuss difference and similarity between these two in our future work.}
, but it's not surprise to have such behaviour in many other models.
It would also be interesting to understand the relation of the boundary information metric to the quantum information theoretic proof of the $g$-theorem as in \cite{Casini:2016fgb,Casini:2018cxg,Casini:2022bsu}.
\subsection{Supersymmetric Case}
Next, we consider a supersymmetric theory:
    \begin{align}
        S_{\textrm{bulk}} \, = \, \frac{1}{2} \int dz d\bar{z} \Big( \pa X \bar{\pa} X + \psi \bar{\pa} \psi + \tilde{\psi} \pa \tilde{\psi} \Big) \, , 
    \end{align}
with the boundary mass perturbation
    \begin{align}
        \pa_\t X + u X \, &= \, 0 \, ,\\
        \pa_\t \psi + u \psi \, &= \, i \h (\pa_\t \tilde{\psi} - u \tilde{\psi} ) \, ,
    \end{align}
on the boundary, where $\h=\pm1$.

\begin{figure}
    \centering
        \includegraphics[width=7.6cm]{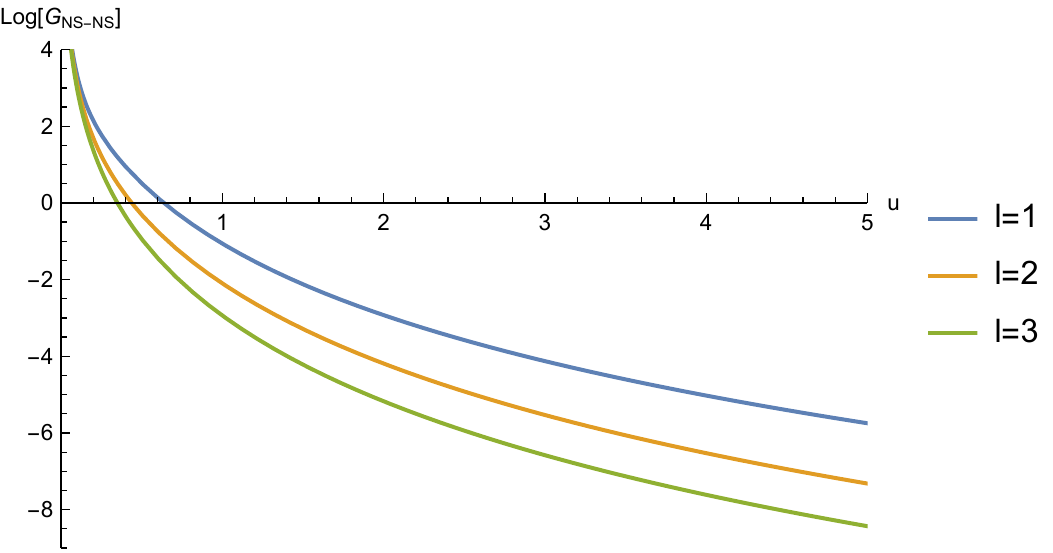} \ \ \includegraphics[width=7.6cm]{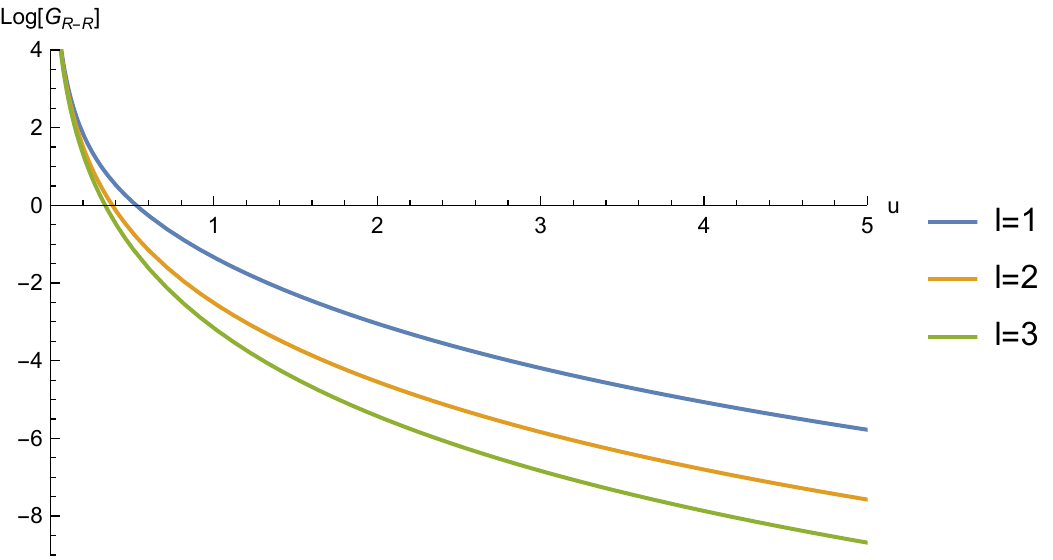}
        \caption{Left: Log plot of the information metric $G$ with $l=1,2,3$ for the NS-NS sector (left) and the R-R sector (right),
        where we chose $n_{\textrm{max}}=10$ and $r_{\textrm{max}}=19/2$.}
\label{fig:GNSNS}
\end{figure}

The cylinder amplitudes are computed in \cite{Arutyunov:2001nz} as
    \begin{align}
        Z_{\textrm{NS-NS}}(u,v) \, &= \, \mathcal{N}_{\textrm{NS-NS}} \, 4^{u+v} u v B(u,u) B(v,v) (u+v+luv)^{-1/2} \nn\\
        &\qquad \times \frac{f_3^7(q)f_3^{(u,v)}(q) - f_4^7(q)f_4^{(u,v)}(q)}{f_1^7(q)f_1^{(u,v)}(q)} \, ,
    \end{align}
for the NS-NS sector, and 
    \begin{align}
        Z_{\textrm{R-R}}(u,v) \, = \mathcal{N}_{\textrm{R-R}} \, \sqrt{\frac{uvq}{u+v+luv}} \frac{f_2^7(q)f_2^{(u,v)}(q)}{f_1^7(q)f_1^{(u,v)}(q)} \, ,
    \end{align}
for the R-R sector. The numerical constants $\mathcal{N}_{\textrm{NS-NS}}$ and $\mathcal{N}_{\textrm{R-R}}$ are independent of the boundary mass $u$ or $v$.
Here we defined $q=e^{-l}$ and
    \begin{align}
        f_1^{(u,v)}(q) \, &= \, q^{\frac{1}{12}} \prod_{n=1}^\inf \left( 1 - \frac{n-u}{n+u} \frac{n-v}{n+v} \, q^{2n} \right) \, , \\
        f_2^{(u,v)}(q) \, &= \, \sqrt{2} q^{\frac{1}{12}} \prod_{n=1}^\inf \left( 1 + \frac{n-u}{n+u} \frac{n-v}{n+v} \, q^{2n} \right) \, , \\
        f_3^{(u,v)}(q) \, &= \, q^{-\frac{1}{24}} \prod_{r=\frac{1}{2}}^\inf \left( 1 + \frac{r-u}{r+u} \frac{r-v}{r+v} \, q^{2r} \right) \, , \\
        f_4^{(u,v)}(q) \, &= \, q^{-\frac{1}{24}} \prod_{r=\frac{1}{2}}^\inf \left( 1 - \frac{r-u}{r+u} \frac{r-v}{r+v} \, q^{2r} \right) \, ,
    \end{align}
with $f_i(q)=f_i^{(0,0)}(q)$.
We show numerical plots for the information metrics computed from these cylinder amplitudes in Figure~\ref{fig:GNSNS}.

\section{Gravity Dual of the Boundary Information Metric}\label{sec5}
In this section, we conjecture a gravity dual of the boundary information metric.
Later, we will see that the gravity dual of the boundary information metric appears as a geometrical object, which we call an ``area of the minimal cross section.''

To clarify the problem, let us consider AdS$_3$/BCFT$_2$ at finite temperature. 
Here, we consider a connected end-of-the-world brane (ETW brane) in the thermal AdS$_3$ background.
The metric of an Euclidean thermal AdS$_3$ is given by
\be
ds^2=R^2\left(\frac{d\tau^2}{z^2}+\frac{dz^2}{h(z)z^2}+\frac{h(z)dx^2}{z^2}\right),
\ee
where $h(z)=1-(z/z_0)^2$. Note that here we compactify the $x$ and $\tau$ as
\begin{align}
    x&\sim x+2\pi z_0,\nn\\
    \tau&\sim\tau+2\pi z_H.
\end{align}
The asymptotic AdS region is given by $z=0$ and we take the region of BCFT as 
\be
0\leq x\leq\pi z_0.
\ee
Next we determine the position of a brane. We can solve the equation of motion for a brane and the profile is solved as \cite{Miyaji:2021ktr,Fujita:2011fp}
\be
x(z)=z_0\arctan\left(\frac{RTz}{z_0\sqrt{h(z)-R^2T^2}}\right)\label{74}
\ee
 for $0\leq x\leq\pi z_0/2$ and 
 \be
 x(z)=\pi z_0-z_0\arctan\left(\frac{RTz}{z_0\sqrt{h(z)-R^2T^2}}\right)\label{75}
 \ee
for $\pi z_0/2\leq x\leq\pi z_0$.

The gravity dual of an information metric between the overlap of states in 2-dim CFT has already been considered in \cite{Miyaji:2015woj}. In that case, the gravity dual of an information metric is expected to represent the area of the time-slice hypersurface in the Janus solution \cite{Bak:2007jm}. Here, we consider the information metric of a boundary counterpart. The overlap of boundary states is always divergent and so is the information metric. Thus, we expect that the gravity dual also contains divergence. In 2-dim CFT the boundary information metric is proportional to an inverse power of modulus in the short length cylinder limit. This can be expected as follows. Let us consider an overlap of boundary states on a cylinder with modulus $s$
\be
Z_{ab}=\bra{a}e^{-sH}\Ket{b},
\ee
where we consider the $s\ll1$ case.
In the open string picture, we can write the partition function as 
\be
Z_{ab}=\operatorname{Tr}e^{-\frac{1}{s}H_{open}}= e^{-\frac{1}{s}(E_0(\lambda_a,\lambda_b)+\cdots)},
\ee
where we expand the energy spectrum in the open string channel and consider the vacuum energy $E_0$. Here, the $\lambda$ is a modulus of the marginal deformation.
In general the energy spectrum $E$ depends on the modulus $s$, but in the short cylinder limit $E_0$ is independent of $s$. In the $s\rightarrow 0$ limit, the $E_0$ takes a finite value and its expansion series start with the constant (as a function of $s$). Therefore, $E_0$ is a function of $\lambda_a$ and $\lambda_b$. By considering the Taylor expansion with respect to the $\lambda_b-\lambda_a$ the information metric reads 
\be
ds^2= \frac{1}{s} d\lambda_i^2+O(s^0),\label{712}
\ee
where $\lambda_i$ are the coordinates of the moduli space. From this, we can argue that the gravity dual of this boundary information metric also behaves as $1/s$. In the thermal AdS$_3$ a natural candidate of this gravity dual is a minimal cross section of the ETW brane. We note that in the BTZ black hole background there is no gravity dual  because the ETW branes are disconnected. This matches with the CFT calculation (\ref{712}), where the BTZ black hole corresponds to an $s\rightarrow0$ or a long cylinder limit and the metric shrinks. Taking into the above facts we conjecture that
 \vskip\baselineskip
  \textit{the gravity dual of the boundary information metric is an area of a cross section of a connected end-of-the-world brane, which minimizes its area.}
\vskip\baselineskip

\begin{figure}
    \begin{center}
    \includegraphics[width=100mm]{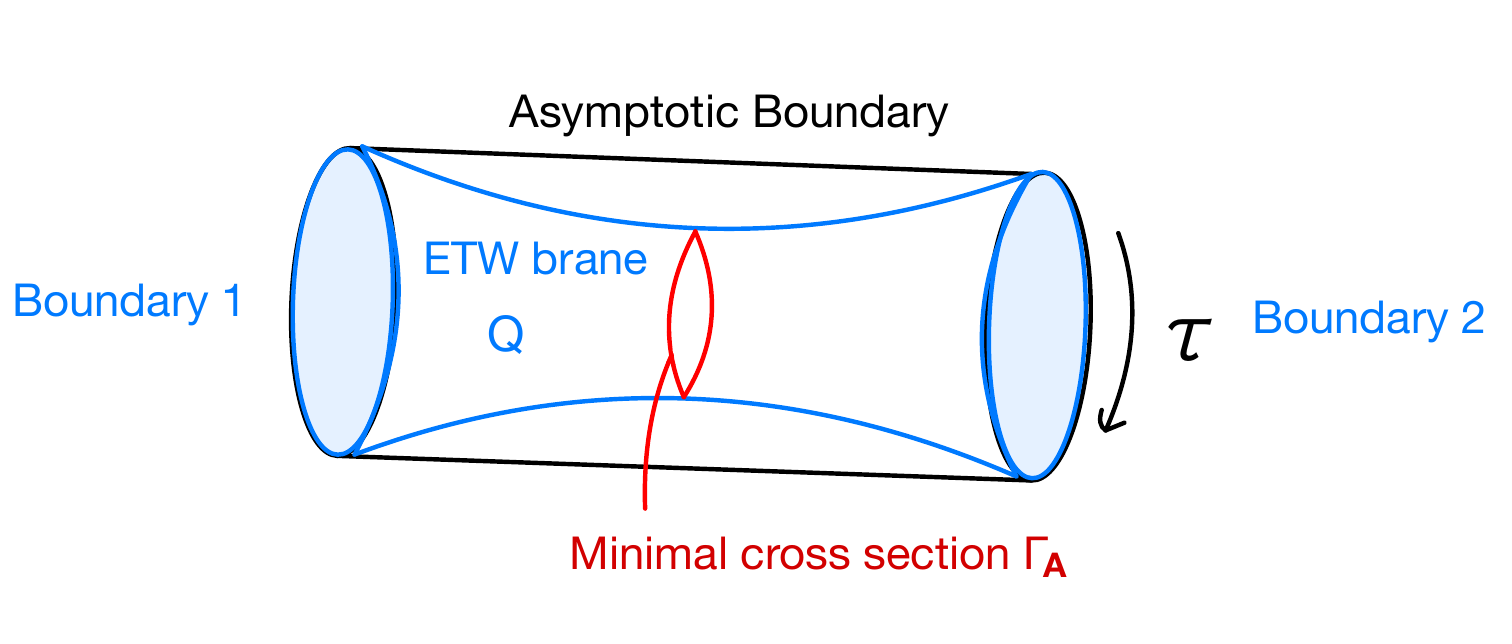}
    \end{center}
    \caption{A schematic figure of a minimal cross section in the thermal AdS$_3$, which is expected to be  dual of the boundary information metric.
    The two BCFTs live on each of the boundaries of the cylinder and boundary states live on the two corners of the cylinder. The minimal cross section $\Gamma_A$ is determined by minimizing the area of the whole cross sections on the ETW brane $Q$.}
\end{figure}

We will call this gravity dual as a ``minimal cross section'' of the ETW brane. Notice that the short length cylinder limit corresponds to the $\beta=2\pi z_0\rightarrow0$ limit. In this limit the minimal cross section of an ETW brane approaches the asymptotic boundary and it indeed behaves as 
\be
G_{\lambda\lambda}\sim \textrm{Area}[\Gamma_A]=\frac{2\pi z_H R}{z_0\sqrt{1-T^2R^2}},\label{79}
\ee
which indeed diverges in a inverse power of a modulus $z_0$. Note that if we make a quess that the gravity dual is another cross section such as $z=const\neq z_0\sqrt{1-R^2T^2}$, then the area of this cross section behaves $1/z$. Of course, we can manipulate the constant slice as $z=az_0$, but we expect that this is not the case. Firstly, as we notice later we have $\mathbf{Z}_2$ symmetry $x\rightarrow\pi R/2-x$ and it is natural to expect that the gravity dual should correspond to the geometrical object at the turning point $x=\pi R/2$. Secondly, as this can also be described just below, the action of the scalar field on the ETW brane has a divergence at the turning point, which means that the integrand at the turning point effectively contributes.

Here, we denote the abstract coordinate of moduli as $\lambda$ and we call the minimal cross section of the ETW brane as $\Gamma_A$. Note that this is the first order approximation of the kink solution as we revisit in the end of this section.
We also note that this conjecture is an explicit example of our main statement:
\vskip\baselineskip
{\it the information metric of boundary states corresponds to the metric of moduli space of a target space}.
\vskip\baselineskip

\begin{figure}
    \centering
        \includegraphics[width=5cm]{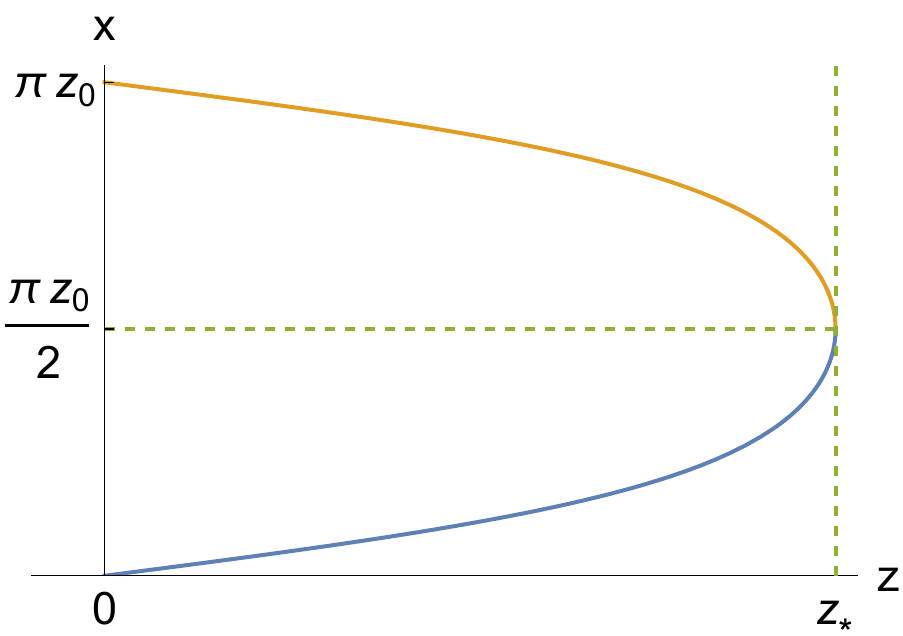} \ \includegraphics[width=5cm]{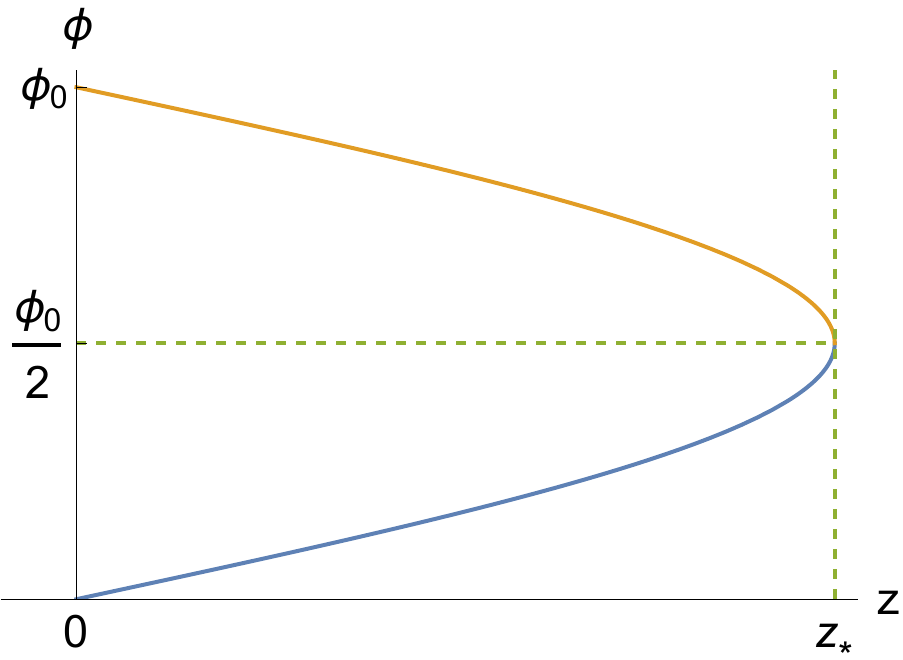} \ \includegraphics[width=5cm]{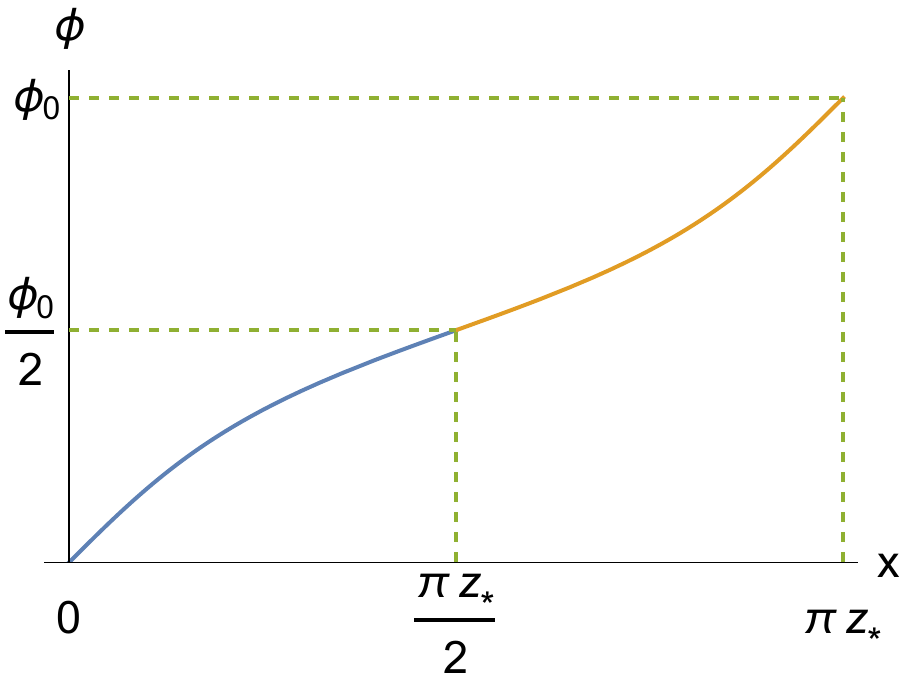}
        \caption{Schematic plots for the relations among $x$, $z$ and $\phi$. The blue region corresponds to $0\leq x \leq\pi z_0/2$ and the orange region corresponds to $\pi z_0/2\leq x \leq\pi z_0$.}
\label{fig:x-z}
\end{figure}

To justify the above conjecture, let us consider a more concrete a example: a real massless scalar field on the ETW brane with the following action
\be
S_{scalar}=\sigma\int_Q d^2x\sqrt{h} (\partial\phi)^2,\label{76}
\ee
where $\sigma$ is some constant.
This corresponds to considering an exactly marginal deformation on the coset coming from breaking $U(1)$ symmetry.
Note also that even though there is also a configuration with disconnected ETW branes as an on-shell solution, its on-shell action is smaller than the one we will consider in the $z_0/z_H\to 0$ limit, as was pointed out in \cite{Miyaji:2021ktr}.
If we solve the equation of motion with boundary conditions $\phi(x=0)=0$ and $\phi(x=\pi z_0)=\phi_0$ under a fixed induced metric $h$,
we expect that there is a sharp jump of $\phi$ around $x=\pi z_0/2$. This can be explained as follows. The induced metric on the brane profile is
evaluated from (\ref{74}) and (\ref{75})
\begin{align}
    h_{\tau\tau}=\frac{R^2}{z^2} \, , \qquad 
    h_{zz}=\frac{R^2}{z^2(h(z)-R^2T^2)}.
\end{align}
The Klein-Gordon equation for the massless field becomes
\be
\Box\phi=\frac{1}{\sqrt{h}}\partial_z(\sqrt{h}h^{zz}\partial_z\phi)=0,
\ee
where we assumed that $\partial_\tau\phi=0$ for rotational invariance. Then, the first derivative can be solved as
\be
\partial_z\phi=\frac{C_1}{\sqrt{h(z)-T^2R^2}},\label{513}
\ee
where $C_1$ is some integration constant. We can integrate again and obtain
\be
\phi=C_1z_0\arctan\left(\frac{z}{z_0\sqrt{h(z)-R^2T^2}}\right)+C_2,\label{514}
\ee
where $C_2$ is another integration constant. The problem is that $z$ is not single valued with respect to $x$ (for example, we have $z=0$ for $x=0$ and $x=\pi z_0$) and this makes us hard to impose the boundary conditions at $z=0$. From (\ref{513}) $\phi$ monotonically increases as $0\leq x\leq \pi z_0$ and taking into account the symmetry $x\leftrightarrow \pi z_0-x$ we find that
\be
\phi(x) = 
\begin{cases}
\phi(z), & (0\leq x \leq\pi z_0/2)\\
\phi_0-\phi(z), & (\pi z_0/2\leq x \leq\pi z_0)\label{515}
\end{cases}
\ee
where we suitably use $x$ and $z$ in each region.
From this, the boundary conditions can be rewritten as 
\begin{align}
    \phi(z=0) \, = \, 0 \, , \qquad 
    \phi(z_*) \, = \, \frac{\phi_0}{2} \, ,
\end{align}
where $z_*=z_0\sqrt{1-R^2T^2}$ represents the turning point of the ETW brane. Substituting this into (\ref{515}) we find that 
\begin{align}
C_1 \, = \, \frac{\phi_0}{\pi z_0} \, , \qquad
C_2 \, = \, 0 \, .
\end{align}
Note that this determines uniquely the $\phi(x)$ via (\ref{515}).
Schematic plots for the relations among $x$, $z$ and $\phi$ are shown in figure~\ref{fig:x-z}.

With these background solutions and the action ($\ref{76}$), we can evaluate the tree-level on-shell action as
\be
S_{scalar}=\frac{2\sigma z_H \phi_0^2}{z_0}
\ee
 and we expect that this represents the gravity dual of a boundary information metric because
  \begin{align}
     \braket{\phi(x=0)=0|\phi(x=\pi z_0)=\phi_0}&=e^{-S_{scalar}}\nn\\
     &\sim1-\frac{2\sigma z_H }{z_0}\phi_0^2.
 \end{align}
 Then, the information metric reads
 \be
 G_{\phi\phi}=\frac{4\sigma z_H}{z_0},\label{scalar}
 \ee
 which reproduces the expected result $G_{\lambda\lambda}\propto 1/z_0$ as in (\ref{79}).
 
Note that the gravity dual of the boundary information metric is ambiguous by mutiplying some $O(1)$ coefficient. For example, in the perturbation of boundary scalar fields the gravity dual has a smooth gradient of scalar fields. The statement that the gravity dual of the information metric is given by the minimal cross-section means that we approximate such a solution with a kink-like profile. 
The situation is the same in \cite{Miyaji:2015woj} where the gravity dual of the bulk information metric was studied.

  Even though we only considered an exact marginal deformation in the direction of the $U(1)$ coset, it would be possible to generalise this to other parameter spaces.
  For exactly marginal boundary deformations, the computation of the information metric corresponds to evaluating the on-shell action with corresponding scalar fields on top of the connecting ETW brane.
  We expect similar computation as what we do in this case.
  Generalisations to relevant and irrelevant deformations are also possible with a suitable potential for the scalar fields on the ETW brane \cite{Kanda:2023zse}.
  It would be interesting to explicitly carry out the computation to study if the monotonicity of the information metric is holographically satisfied or not.

\section{Conclusion}
In this paper, we introduced a novel information metric whose distance is defined by the inner product of boundary states, which we called a boundary information metric. We expect that this boundary information metric describes a metric of a moduli space in a target space under a suitable perturbation of boundary states and confirmed this by explicit calculations. We examined the boundary information metric in the SU(2) WZW model and a mass deformed real free scalar and its SUSY counterpart in 2-dim CFT. In the SU(2) WZW model, we can reproduce the metric of the target space, a metric of the SU(2).  In the mass deformed free scalar case, under a flow from the Neumann boundary condition to the Dirichlet boundary condition, the boundary information metric is monotonically decreasing, although the disk amplitude itself is not. In the supersymmetric case, the information metric monotonically decreases. However, the action (\ref{S}) defined from the disk amplitude  becomes also monotonic both in the bosonic and supersymmetric cases.

     As an application to the holography, we conjectured that the boundary information metric corresponds to the area of a minimal cross section of an ETW brane in the thermal AdS space, which corresponds to a short cylinder limit. We confirmed this conjecture by calculating the area of the minimal cross section and reproduce a modulus dependence $1/z_0$. To confirm more concretely, we also considered a real free scalar field on the ETW brane and calculated the action with perturbed boundary conditions. From this, we evaluated the boundary information metric and reproduced the modulus dependence $1/z_0$, similarly. In this calculation, we assumed that the gradient of the fields on the ETW brane changes so sharply that it localizes as a minimal cross section geometrically in the first order approximation. Therefore, we expect that it is hard to reproduce the coefficients of the boundary information metric from the calculation of the minimal cross section on the ETW brane.  Toward explicit formulation of the boundary states in AdS/BCFT, symmetric orbifold formulation seems to be useful, which will be one of future directions of our boundary information metric \cite{Eberhardt:2018ouy,Belin:2021nck,Takayanagi:2022xpv}. We can also examine the behavior of the information metric under boundary RG-flows including the relevant deformation. We can consider scalar field on the ETW brane and tune the potential as in \cite{Kanda:2023zse}. We expect that we can analyze the information metric under the backreacted solution.
     
 Finally, we briefly commented on the relation to the string theory in the Appendix \ref{sec:d-branes}. In this analysis, we can reproduce the inverse power of the modulus dependence for the fixed modulus and the flat metric of the transverse directions to the D-branes. We cannot interpret this metric in string theory directly because we do not integrate over the moduli of the worldsheet, but even in that case, we can probe the metric of the target space even though its coefficient seems to be meaningless. It will be a future direction to study the relation of the this boundary information metric calculated with the modulus fixed and the target space geometry.

\section*{Acknowledgements}
We are grateful to Tadashi Takayanagi for stimulating discussions and support throughout the project.
This work is supported by MEXT KAKENHI Grant-in-Aid for Transformative Research Areas (A) through the ``Extreme Universe'' collaboration: Grant Number 21H05187.
The work of KS is also supported by the Simons Foundation through the ``It from Qubit'' collaboration.
MW is supported by Grant-in-Aid for JSPS Fellows (No. 22J00752).

\appendix
\section{Boundary Information Metric in Superstring}
\label{sec:d-branes}
 In this section, we will consider boundary states in a worldsheet CFT of superstring theory and check that the boundary information metric  corresponds to the metric in the target space, which is just a flat 10-dim Minkowski space. Below, we will follow the argument in \cite{Polchinski:1996fm,DiVecchia:1999mal}.

\begin{figure}
 \begin{center}
  \includegraphics[width=100mm]{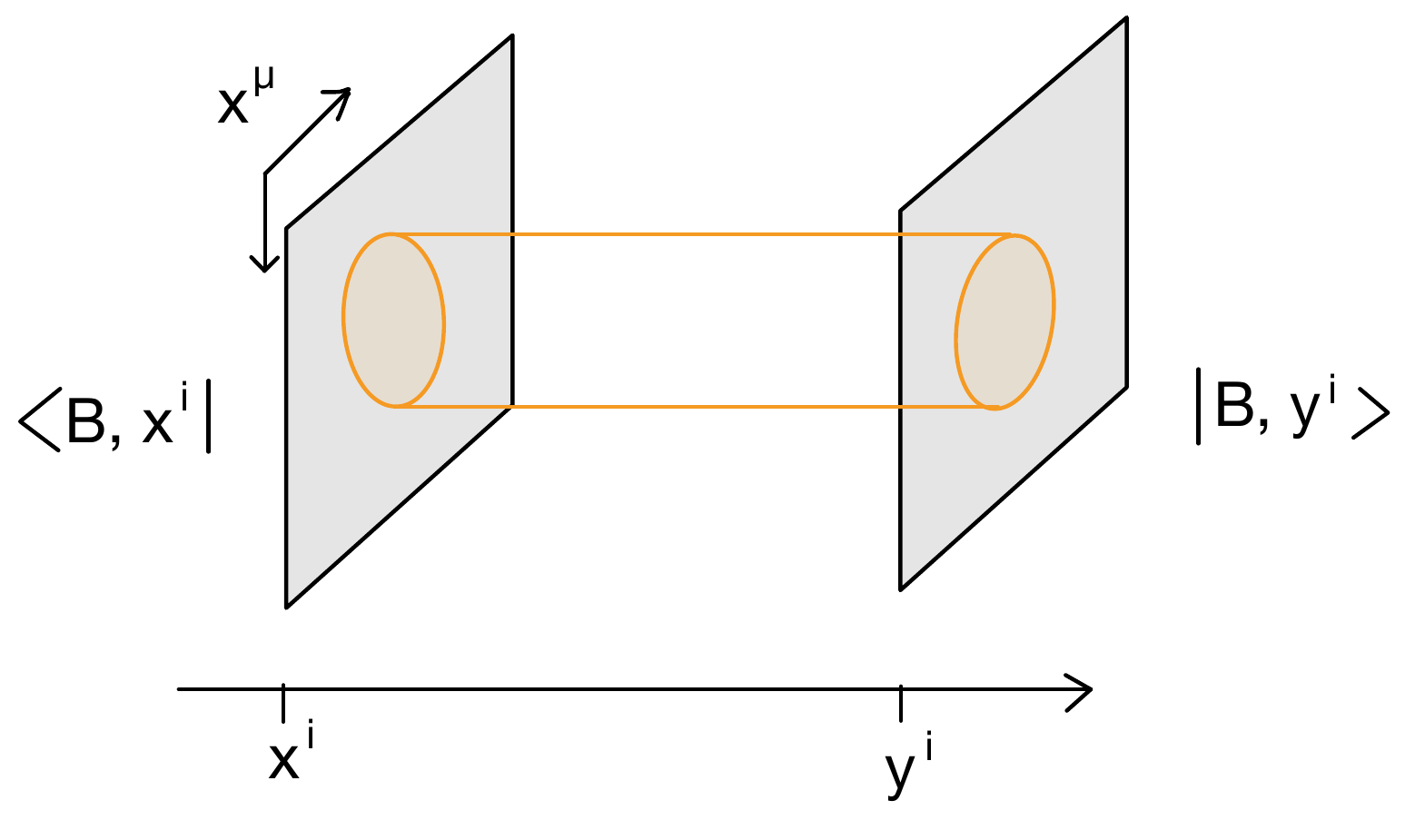}
 \end{center}
 \caption{A schematic figure of D-branes scattering. Two parallel D-branes exchange closed strings and generate forces coming from a RR-sector and a NSNS-sector. Usually, we evaluate the amplitude and fix the tension of D-branes.} 
 \label{Fig1}
\end{figure}

We consider a Dp-brane extending along the $x^{0},\cdots,x^{p}$ directions. Then, the boundary conditions of open strings can be described as 
\begin{align}
    &\partial_{\tau}X^{\mu}\Ket{B,\eta}=(\psi^{\mu}+i\eta\tilde{\psi^{\mu}})\Ket{B,\eta}=0,\nn\\
    &\partial_{\sigma}X^{i}\Ket{B,\eta}=(\psi^{i}-i\eta\tilde{\psi^{i}})\Ket{B,\eta}=0,
\end{align}
where we represent the tangential and perpendicular directions to the D-brane as $x^{\mu}$ and $x^i$, respectively. The $\eta$ labels the spin structure of D-branes or the R and NS sector of open strings stretching between D-branes. In terms of the creation and annihilation operators we can rewrite these boundary conditions as 
\begin{align}
    &(\alpha^{\mu}_{n} +\tilde{\alpha}^\mu_{-n})\Ket{B,\eta}=0,\nn\\
    &(\alpha^{i}_{n} -\tilde{\alpha}^i_{-n})\Ket{B,\eta}=0,\nn\\
    &(\psi^{\mu}_r+i\eta\tilde{\psi}^{\mu}_{-r})\Ket{B,\eta}=0,\nn\\
     &(\psi^{i}_r-i\eta\tilde{\psi}^{i}_{-r})\Ket{B,\eta}=0.
    \end{align}
We can integrate these and the boundary states can be described by coherent states
\be
\Ket{B,\eta}\sim\exp{\sum_{n=0}^\infty\frac{1}{n}(\ap^\mu_{-n}\tilde{\ap}^\mu_{-n}-\ap^i_{-n}\tilde{\ap}^i_{-n})-i\eta\sum_{r>0}^\infty}(\psi^\mu_{-r}\tilde{\psi}^\mu_{-r}-\psi^i_{-r}\tilde{\psi}^i_{-r})\Ket{B,\eta}_{(0)},
\ee
where the $\Ket{B,\eta}_{(0)}$ represents a boundary state for zeromodes. To move onto zeromodes let us firstly consider the NS sector. We will denote the left and right  worldsheet fermion number as $F$ and $\tilde{F}$, respectively. Their actions on the boundary states are
\begin{align}
   & (-1)^F\Ket{B,\eta}_{NSNS}=-\Ket{B,-\eta}_{NSNS},\nn\\
   &(-1)^{\tilde{F}}\Ket{B,\eta}_{NSNS}=-\Ket{B,-\eta}_{NSNS}.
\end{align}
 Then, GSO invariant boundary states are obtained as
 \be
\Ket{B}_{NSNS}=\frac{1+(-1)^F}{2}\frac{1+(-1)^{\tilde{F}}}{2}\Ket{B,+}_{NSNS}=\frac{1}{2}(\Ket{B,+}_{NSNS}-\Ket{B,-}_{NSNS}).
\ee
In the RR sector we can similarly obtain
\be
\Ket{B}_{RR}=\frac{1+(-1)^{F+p}}{2}\frac{1+(-1)^{\tilde{F}+p}}{2}\Ket{B,+}_{RR}=\frac{1}{2}(\Ket{B,+}_{RR}+\Ket{B,-}_{RR}).
\ee
From this we can calculate an amplitude between two parallel Dp-branes at $x^i$ and $y^i$ in the long length approximation ($s\rightarrow\infty$) as in (Fig.\ref{Fig1}) \cite{Polchinski:1996fm}
\begin{align}
\mathcal{A}_{x^iy^i}&=\frac{\alpha'}{2}\frac{T_p^2}{4}V_{p+1}\int ds(\frac{dk}{2\pi})^{9-p} e^{ik(x-y)-\frac{s}{2}\alpha'k^2}(\frac{f_3(q)^8-f_4(q)^8-f_2(q)^8}{2f_1(q)^8}) \label{47} \\
&=\frac{\alpha'T_p^2}{16}V_{p+1}\int \frac{\pi dt}{t^6}(\frac{t}{2\pi^2\alpha'})^{\frac{9-p}{2}} e^{-\frac{(x-y)^2}{2\pi\alpha'}t}(\frac{f_3(\tilde{q})^8-f_4(\tilde{q})^8-f_2(\tilde{q})^8}{2f_1(\tilde{q})^8}) \label{48} \\
&\sim(1-1)V_{p+1}2\pi(4\pi^2\ap')^{3-p}G_{9-p}(|x^i-y^i|^2) \label{49},
\end{align}
where $G_D(y)=\frac{\Gamma(\frac{D}{2})}{2\pi^{\frac{D}{2}}}y^{2-D}$ is a massless scalar Green's function in $d$ dimensions and $q$ and $\tilde{q}$ are given by
\begin{align}
    q&=e^{-s}\nn\\
    \tilde{q}&=e^{-\pi t}=e^{-\frac{\pi^2}{s}},
\end{align}
where $s$ is the modulus of a cylinder in the closed string channel. However, here we need a short cylinder limit ($s\rightarrow0$). In this limit the amplitude can be approximated as
\begin{align}
    \mathcal{A}_{x^iy^i}&=\frac{\alpha'}{2}\frac{T_p^2}{4}V_{p+1}\int ds\left(\frac{dk}{2\pi}\right)^{9-p} e^{ik(x-y)-\frac{s}{2}\alpha'k^2}\left(\frac{f_3(q)^8-f_4(q)^8-f_2(q)^8}{2f_1(q)^8}\right) \nn\\
    &=\frac{\alpha'}{16}\frac{T_p^2}{(2\pi^3\al)^{\frac{9-p}{2}}}V_{p+1}\int_0^\infty dss^{\frac{p-1}{2}}e^{-\frac{(x-y)^2}{2\al s}}\left(\frac{f_3(q)^8-f_4(q)^8-f_2(q)^8}{f_1(q)^8}\right)\label{311}
\end{align}

In the calculation of 2-dim CFT, we usually introduce a cutoff for the boundary state via
 \be
 \Ket{\psi}=e^{-\ep H}\Ket{B}.
 \ee
Keeping this in mind, it is natural to treat the modulus as a cutoff because here the $\ep$ corresponds to the cylinder modulus $s$. Note that this cut-off is not compatible with the calculation in the string theory because we do not integrate over the modulus.  For the fixed modulus, we can calculate the boundary information metric in the $y\rightarrow x$ limit from (\ref{311}) 
\begin{align}
D_B^2
     &=(1-1)\frac{2}{\pi\alpha's}dx_i^2,\label{713}
     \end{align}
which reproduces the flat metric along the transverse direction to the D-branes. Here, we again note that we do not integrate over the modulus of a cylinder. In the $s\rightarrow\infty$ limit the boundary information metric approaches to zero. If we consider $s\rightarrow0$ limit or a short distance limit, it indeed diverges as $1/\ep$. Note that if we consider a long cylinder limit for the fixed modulus, then the information metric is the same as (\ref{713}).  

Several comments are in order. Firstly, the cylinder amplitude indeed vanishes due to the BPS condition (we denote as $(1-1)$ in front of (\ref{49})). One may wonder if it is meaningless to take the variation of the originally vanishing quantity. However, it does not affect our statement because we can only consider the RR sector or NSNS sector individually. In the NSNS sector the exchange of gravitons and dilatons generates the attraction force and in the RR sector the exchange of the RR fields generates the repulsion force. Because the boundary states are proportional to the D-brane tension, the magnitude of the amplitude is also proportional to the tension and has physical meanings.

Secondly, we note that the boundary information metric depends on the modulus. In the long cylinder limit in the closed string channel, where $s\rightarrow\infty$, the boundary information metric becomes zero. In this limit, only the massless modes survive. This is consistent with a physical intuition. If we change the position of branes in the long cylinder limit, we can only detect the massless modes and the disk amplitudes. In the long cylinder limit, the energy spectrum always takes the vacuum. The translation along the transverse direction does not change the tension of the branes, hence the information metric vanishes. We can  generalize this logic explicitly below.

 Now, let us consider an overlap of boundary states with boundary conditions $a$ and $b$, which is slightly perturbed from $a$ by a relevant operator $\mathcal{O}$
 \be
S\rightarrow S+\lambda\int dx \mathcal{O}.\label{414}
\ee
Especially, here we consider an almost marginal operator for simplicity.
In the long cylinder limit, the overlap of boundary states factorizes into the disk amplitude and the bulk propagator
\be
\braket{a|b}\sim g_a g_be^{-E_0L},
\ee
where we define the disk amplitude $g_a=\braket{a|0}$. In the above, we also approximate the energy of intermediate particles by the vacuum one. In this limit, the boundary information metric can be approximated as
\begin{align}
    D_B^2=&2\left(1-\abs{\frac{\braket{a|b}}{\sqrt{\braket{a|a}}\sqrt{\braket{b|b}}}}\right)\nn\\
    &= 2\left(1-\abs{\frac{g_a g_b e^{-E_0L}}{\sqrt{g_a^2 e^{-E_0 L}}\sqrt{g_b^2 e^{-E_0 L}}}}\right)\cdots\nn\\
    &=0,
\end{align}
where we ignore the sub-leading terms since they are exponentially suppressed $\mathcal{O}(e^{-E_n L})$ and $E_n$ is the energy of the n-th excited state.

The inverse power of modulus dependence in the boundary information metric is a basic feature of 2-dim CFTs. These results match with the ones in the AdS$_3$/BCFT$_2$ case. Especially, the boundary information metric becomes divergent (an inverse power of the modulus) in the short length limit of the cylinder (thermal AdS$_3$) and zero in the long cylinder limit (the BTZ black-hole).

\bibliographystyle{JHEP}
\bibliography{gegeupload.bib}

\end{document}